\documentclass{aastex}
\usepackage{emulateapj5}
\usepackage{apjfonts}

\newcommand{\rt}{\tilde{r}}
\newcommand{\lt}{\tilde{l}}
\renewcommand{\min}{{\rm min}}
\renewcommand{\max}{{\rm max}}
\newcommand{\disk}{{\rm disk}}

\newcommand{\pfrac}[2]{{\left(\frac{#1}{#2}\right)}}

\newcommand\Mdot{{\dot{M}}}
\newcommand\Msun{{M$_\odot$}}
\newcommand\rmin{{r_{\rm min}}}

\slugcomment{Accepted for publication in the Astrophysical Journal}

\begin{document}

\title{An X-ray Reprocessing Model of Disk Thermal Emission in\\
Type 1 Seyfert Galaxies}

\author{James Chiang\altaffilmark{1}} 

\affil{NASA/GSFC, Code 661, Greenbelt MD 20771} 
\altaffiltext{1}{Also at Joint Center for Astrophysics/Physics Department, 
         University of Maryland, Baltimore County, Baltimore MD 21250}

\begin{abstract}
Using a geometry consisting of a hot central Comptonizing plasma
surrounded by a thin accretion disk, we model the optical through hard
X-ray spectral energy distributions of the type 1 Seyfert galaxies NGC
3516 and NGC 7469.  As in the model proposed by Poutanen, Krolik, \&
Ryde for the X-ray binary Cygnus X-1 and later applied to Seyfert
galaxies by Zdziarski, Lubi\'nski, \& Smith, feedback between the
radiation reprocessed by the disk and the thermal Comptonization
emission from the hot central plasma plays a pivotal role in
determining the X-ray spectrum, and as we show, the optical and
ultraviolet spectra as well.  Seemingly uncorrelated optical/UV and
X-ray light curves, similar to those which have been observed from
these objects can, in principle, be explained by variations in the
size, shape, and temperature of the Comptonizing plasma.  Furthermore,
by positing a disk mass accretion rate which satisfies a condition for
global energy balance between the thermal Comptonization luminosity
and the power available from accretion, one can predict the spectral
properties of the heretofore poorly measured hard X-ray continuum
above $\sim 50$\,keV in type 1 Seyfert galaxies.  Conversely,
forthcoming measurements of the hard X-ray continuum by more sensitive
hard X-ray and soft $\gamma$-ray telescopes, such as those aboard the
{\em International Gamma-Ray Astrophysics Laboratory} ({\em INTEGRAL})
in conjunction with simultaneous optical, UV, and soft X-ray
monitoring, will allow the mass accretion rates to be directly
constrained for these sources in the context of this model.
\end{abstract}

\keywords{galaxies: active --- galaxies: individual (NGC 3516, NGC 7469) 
--- galaxies: Seyfert --- X-rays: galaxies}

\section{Introduction}
An important characteristic of type 1 Seyfert galaxies (hereafter
Seyfert 1s) is the highly correlated variability of the continuum flux
across a wide range of optical and ultraviolet wavelengths.  The short
relative lags between the various bands and the observed optical/UV
color variations have been interpreted as evidence of thermal
reprocessing by the accretion disk of higher energy emission from a
single, relatively compact source (Krolik et al.\ 1991; Courvoisier \&
Clavel 1991; Collin-Souffrin 1991).  The principal candidate for this
emission is the power-law-like X-ray continuum which is believed to be
produced via thermal Comptonization in regions of hot plasma near the
central black hole.  Observations of relativistically broadened iron
K$\alpha$ emission lines and the so-called Compton reflection hump
provide further evidence that a substantial amount of X-ray flux is
intercepted and reprocessed by a thin accretion disk in these systems.

Recent multiwaveband monitoring observations of Seyfert 1s have shown
that the optical/UV emission is not related to the observed X-ray flux
in any simple way.  In June--July 1996, a $\sim 30$ day UV and X-ray
monitoring campaign of NGC 7469 by the {\em International Ultraviolet
Explorer (IUE)} and the {\em Rossi X-ray Timing Experiment (RXTE)}
showed that although the UV and 2--10 keV X-ray light curves displayed
some rough similarities in variability amplitudes and time scales,
they were not related in a fashion that would naively be expected from
thermal reprocessing (Nandra et al.\ 1998).  In particular, the UV
maxima preceded similar features in the X-rays by $\sim 4$ days, while
the minima in both light curves occurred nearly simultaneously.  A
more recent analysis of these data revealed, however, that the UV
light curve is well correlated with the X-ray spectral index as well
as with the extrapolated X-ray flux in the 0.1--100\,keV band.  This
suggests that there is a causal connection between the X-ray and UV
emission in this object (Nandra et al.\ 2000).  In April 1998, an
intensive 3-day optical/UV/X-ray monitoring observation of NGC 3516 by
the {\em Hubble Space Telescope (HST)} and {\em RXTE} showed optical
fluxes which changed only by $\sim 3$\% (peak-to-peak) while the
2--10\,keV X-ray flux varied by $\sim 60$\%.  Furthermore, the shapes
of the variations in the optical and X-ray light curves were only
marginally similar.  Nonetheless, the variability at optical
wavelengths, from 3590\,\AA\ to 5510\,\AA, was highly correlated with
essentially zero lag between the different continuum bands, consistent
with the thermal reprocessing model (Edelson et al.\ 2000).

Chiang \& Blaes (2001a; hereafter Paper I) have shown that the 1998
optical and X-ray observations of NGC 3516 can largely be explained by
a thermal Comptonization/disk reprocessing model similar to that
proposed by Poutanen, Krolik, \& Ryde (1997) to account for the
differences between the hard and soft spectral states in the Galactic
X-ray binary Cygnus X-1.  This model was also applied in a simplified
form by Zdziarski, Lubi\'nski, \& Smith (1999; hereafter ZLS) to both
X-ray binaries and Seyfert galaxies in order to explain an apparent
correlation between the X-ray spectral index and the relative
magnitude of the Compton reflection component.  In Paper I, we found
that inversely-related changes in the size and X-ray luminosity of the
Comptonizing plasma can produce the relatively large variations in the
2--10~keV band while causing only much smaller variations in the disk
thermal reprocessed emission.  With the addition of disk emission due
to internal viscous dissipation, we found that the optical and X-ray
variations could be roughly matched while simultaneously satisfying an
energy balance condition relating the bolometric luminosity to the
radiative power available through disk accretion.

In this paper, we refine the calculations for NGC 3516 presented in
Paper I by explicitly fitting the optical and X-ray continuum data and
producing broad band spectral energy distributions (SEDs) for each
observing epoch.  In addition, we apply these calculations to the 1996
UV/X-ray monitoring observations of NGC 7469 in order to provide a
more complete physical picture for the correlations found by Nandra et
al.\ (2000).  We proceed by discussing the relevant optical, UV, and
X-ray observations in \S2.  In \S3, we describe the thermal
Comptonization and disk reprocessing model including our modifications
of the geometry of the Comptonizing region and empirical color
corrections for the disk emission.  The results for NGC 3516 and NGC
7469 are presented in \S4; and we discuss these results and conclude
in \S5.

\section{Observations}

For the 2--10 keV fluxes and spectral indices of NGC 3516, we have fit
the archival {\em RXTE/Pro\-portional Counter Array (PCA)} data for
the April 1998 observations.  Our X-ray spectral model consisted of an
exponentially cut-off power-law plus a Compton reflection component
(the {\sc pexrav} model in {\sc xspec v10.0}), a Gaussian line for the
Fe K$\alpha$ emission line and neutral absorption for the foreground
column density.  The cut-off energy for the exponential roll-over was
set at $E_c = 120$\,keV which is sufficiently high so that it does not
affect the inferred power-law index in the 2--10\,keV band.  We find
somewhat softer spectral indices than Nandra et al.\ (1999) who
analyzed the simultaneous observations by the {\em Advanced Satellite
for Cosmology and Astrophysics (ASCA)}.  This discrepancy is due to
the fact that these authors did not include a Compton reflection
component in their analysis (K.\ Nandra 2001, private communication).
The optical continuum variations have been taken from Fig.~2 of
Edelson et al.\ (2000), and the mean optical fluxes at 5510\,\AA,
4235\,\AA, and 3590\,\AA\ are from their Table~1.\footnote{For the
spectral fitting, optical and UV wavelengths have been de-redshifted
to the source rest frame using $z = 0.0088$ and $z = 0.0163$ for NGC
3516 and NGC 7469, respectively.}  Because of the small aperture of
the {\em HST/Space Telescope Imaging Spectrograph}, we do not apply a
starlight correction.  We do make an extinction correction using
E(B$-$V) $= 0.042$ (Schlegel, Finkbeiner, \& Davis 1998) and the
reddening law of Cardelli et al.\ (1989).  The flux measurements at
1360\,\AA\ made by Edelson et al.\ were not simultaneous with the
optical data, and as these authors discuss, the UV measurements are
affected by detector gain changes induced by thermal variations.
Therefore, we restrict our analysis of the disk thermal emission to
the optical bands.  Since spectral fitting of the X-ray data entailed
integration times of $\sim 30$--70\,ks (see Table~\ref{3516_data}),
the optical fluxes used in the SED fits were averaged over the
duration of each of these epochs.  If the optical emission is due to
disk reprocessing, it is expected to be produced at relatively large
distances ($r > 10^{14}$\,cm) from the central black hole.  In accord
with this, we have shifted the optical light curves by the measured
peak lag of $-0.21$ days relative to the 2--10 X-ray light curve
(Edelson et al.\ 2000).  The times of each observing epoch, the X-ray
fit parameters, and the corrected and time-shifted optical fluxes are
given in Table~\ref{3516_data}.

The optical and UV continuum fluxes for NGC 7469 were obtained from
the AGN Watch web page ({\tt
http://astronomy.ohio-state.edu/$\sim$agnwatch/}). We use the
4865\,\AA\ and 6962\,\AA\ continuum light curves measured by Collier
et al.\ (1998) and the modified 1315\,\AA\ light curves of Kriss et
al.\ (2000).  Welsh et al.\ (1998) compared the ground-based optical
spectra of Collier et al.\ to contemporaneous {\em HST} spectra and
found that the ground-based spectra are substantially redder due to
contamination from starburst activity in the host galaxy.  In order to
determine the continuum emission due to the Seyfert nucleus, we used
the difference between the {\em HST} and the mean ground-based spectra
to estimate the starlight contribution.  The simultaneous UV data were
obtained using {\em IUE} (Wanders et al.\ 1997; Kriss et al.\ 2000),
and Welsh et al.\ (1998) showed that the {\em IUE} and {\em HST}
fluxes at 1315\,\AA\ were consistent over a half-day period during the
30-day monitoring campaign. Thus, despite the relatively large
aperture used for the {\em IUE} spectra, we assume that the starlight
contribution to be negligible at this wavelength.  A reddening
correction using E(B$-$V) $= 0.12$ has been applied to the optical and
UV fluxes. This correction includes a Galactic component as well as
absorption intrinsic to NGC 7469 (Kriss et al.\ 2000).  The mean
corrected optical and UV fluxes are $F_{1315} = 11.4$, $F_{4865} =
1.59$, and $F_{6962} = 0.75$, all in units of
$10^{-14}$erg\,cm$^{-2}$s$^{-1}$\AA$^{-1}$.  The X-ray fluxes and
spectral indices are taken from Nandra et al.\ (2000) who fit the {\em
RXTE/PCA} data with the same model we used for NGC 3516.  The optical and
UV fluxes have been linearly interpolated from the AGN Watch data at
the observing times of the X-ray data.  The fluxes at 4865\,\AA\ and
6962\,\AA\ were time-shifted by the measured lags relative to the
1315\,\AA\ light curve: $-1.0$ days for the 4865\,\AA\ data and $-1.5$
days for the 6962\,\AA\ data (Collier et al.\ 1998).

\section{The Model: Thermal Comptonization Geometry, Energy Balance,
and the Disk Spectrum}

The geometry we consider is similar to that proposed by Shapiro,
Lightman, \& Eardley (1976) in their two-temperature disk model of
Cygnus X-1: a central, geometrically thick region of hot plasma
surrounded by a relatively cold, thin accretion disk.  In principle,
the hot central region could arise because of a thermal instability as
originally envisioned by Shapiro et al.\ (1976), or it could consist
of an advection dominated accretion flow (ADAF) (e.g., Narayan \& Yi
1995) or any one of its more recent variations (Blandford \& Begelman
1999; Narayan, Igumenshchev, \& Abramowicz 2000; Quataert \& Gruzinov
2000).  Following Poutanen et al.\ (1997), our goal is to infer the
properties of the disk and Comptonizing plasma from the radiation
physics and thereby provide constraints for the dynamical models of
the inner accretion flow which may help distinguish between the
various candidates.

A key feature of the present calculations is the thermostatic feedback
effect between the X-ray emission from the Comptonizing plasma and the
soft photons which are produced by thermal reprocessing of these
X-rays by the accretion disk (Haardt \& Maraschi 1991; Stern et al.\
1995).  ZLS noted that the degree of feedback can be adjusted by
altering the geometry of the hot plasma, and they invoked this
mechanism to explain the apparent correlation between the X-ray
spectral index, $\Gamma$, and the relative magnitude, $R$, of the
Compton reflection component in Seyfert galaxies and X-ray binaries.
Following Poutanen et al.\ 1997 and similar results Zdziarski et al.\
(1998) obtained by applying a similar analysis to the X-ray binary
GX339$-$4, ZLS proposed that the region of Comptonizing plasma takes
the form of a uniform, optically thin sphere centered on the black
hole.  The X-ray spectral index variations occur because the disk
penetrates the plasma sphere by different amounts due to changes in
the sphere radius, $r_s$, and/or changes in the disk inner truncation
radius, $\rmin$.  Greater overlap of the disk by the sphere results in
a larger fraction of the X-ray emission being intercepted by the disk
and also in a larger amount of reprocessed flux.  This produces an
increased Compton reflection component as well as more soft disk
photons entering the hot plasma.  The additional disk flux provides
enhanced cooling of the plasma by increasing Compton losses; and the
equilibrium temperature of the plasma decreases, thus producing a
softer X-ray spectrum.  Using estimates for the fraction of thermally
reprocessed emission intercepted by the plasma sphere as a function of
the ratio $\rmin/r_s$ and an empirical relationship between the
Compton amplification factor, $A$, and the spectral index of the
thermal Comptonization continuum (Beloborodov 1999),
\begin{equation}
\Gamma = 2.33(A - 1)^{-1/10},
\label{Gamma(A)_old}
\end{equation}
ZLS derived an $R$-$\Gamma$ relation which is similar to the observed
correlation found in the X-ray spectra of Seyfert 1s.

The radial distribution of the flux incident upon a razor-thin disk
due to isotropic emission from an optically thin sphere is given by
\begin{equation}
F_{\rm inc}(r) = \frac{3}{16\pi^2}\frac{L_x}{r_s^2} h\pfrac{r}{r_s},
\label{Finc_sphere}
\end{equation}
where $L_x$ is the total thermal Comptonization luminosity.  The
dimensionless function $h(r/r_s)$ is Eq.~1 of ZLS.  It has the
asymptotic behavior $h(r/r_s \gg 1) \rightarrow (\pi/4)(r_s/r)^3$ and
is approximately flat for $r/r_s \la 1$ with $h(0) = \pi$.  In Paper
I, we calculated the disk thermal spectrum by assuming local blackbody
emission and estimated the local disk temperature as being due to a
combination of reprocessed flux and emission due to internal viscous
dissipation:
\begin{equation}
T_\disk(r) = \left[\frac{(1-a)F_{\rm inc}(r) 
      + F_{\rm visc}(r)}{\sigma_B}\right]^{1/4},
\label{T(r)}
\end{equation}
where $a = 0.15$ is the assumed disk albedo and $\sigma_B$ is the
Stefan-Boltzmann constant.  As in Paper I, here we model the flux due
to viscous dissipation as
\begin{equation}
F_{\rm visc}(r) = \frac{3}{8\pi}\frac{GM\Mdot}{r^3}
                  \left[1 - \pfrac{r_I}{r}^{1/2}\right],
\label{Fvisc(r)}
\end{equation}
where $M$ is the black hole mass, $\Mdot$ is the mass accretion rate,
and $r_I$ is the ``effective'' disk inner radius which we hereafter
take to be $r_I = r_g \equiv GM/c^2$, appropriate for a maximal Kerr
metric.

In Paper I, we showed that the near constant optical continuum of NGC
3516 in the presence of large variations in the X-ray flux could be
explained in this model if the thermal Comptonization luminosity were
inversely proportional to the radius of the plasma sphere, $L_x
\propto r_s^{-1}$.  In addition, we found that the disk inner
truncation radius had to be fairly constant on time scales longer than
the 3-day observation in order to avoid excessive color variations in
the optical and UV bands.  However, if one assumes that the
contribution to the disk thermal emission from internal viscous
dissipation is negligible, then the amount of X-ray luminosity
required to produce the observed optical continuum levels result in
2--10\,keV fluxes which exceed the observed values by a factor $\sim
4$.  The addition of viscous dissipation in the disk can reduce the
needed X-ray luminosity substantially; and as it turns out, the amount
of viscous dissipation required to reproduce the mean X-ray spectrum
also satisfies an energy balance condition:
\begin{equation}
L_{\rm untr} \approx L_x + L_\disk
\end{equation}
where $L_\disk$ is the luminosity of the disk thermal emission
due to viscous dissipation assuming that the inner disk is
cut-off at the truncation radius $\rmin$, and $L_{\rm untr}$ is the
total luminosity from viscous dissipation for an untruncated thin disk
extending all the way down to $r_I$:
\begin{equation}
L_{\rm disk, untr} = 4\pi\int_{\rmin, r_I}^\infty r\,dr\,F_{\rm visc}(r).
   \label{Ldiskuntr}\\ 
\end{equation}
The difference $L_{\rm untr} - L_\disk$ can be interpreted as the
accretion power that would otherwise emerge as local blackbody
emission from the inner regions of the disk that is instead being used
to heat the Comptonizing plasma.

If one neglects local viscous dissipation, the intercepted seed photon
luminosity is given by $L_s = L_x \tilde{L}_s$, where $\tilde{L}_s$ is
a function only of the quantity $\rmin/r_s$ (ZLS).  Since the Compton
amplification factor is the ratio of the thermal Comptonization
luminosity and the seed photon luminosity, $A \equiv L_x/L_s =
1/\tilde{L}_s$, it too is a function only of $\rmin/r_s$.  In
particular, its value monotonically decreases as $\rmin/r_s$
increases.  For $\rmin/r_s = 1$ and using Eq.~\ref{Gamma(A)_old}, the
equilibrium photon index is $\Gamma = 1.55$.  Since the fitted X-ray
spectral indices for NGC 3516 are $\Gamma \sim 1.65$, this implies
that the plasma sphere must overlap disk so that $r_s > \rmin$ for
this model, assuming zero viscous dissipation.

\begin{figure}
\plotone{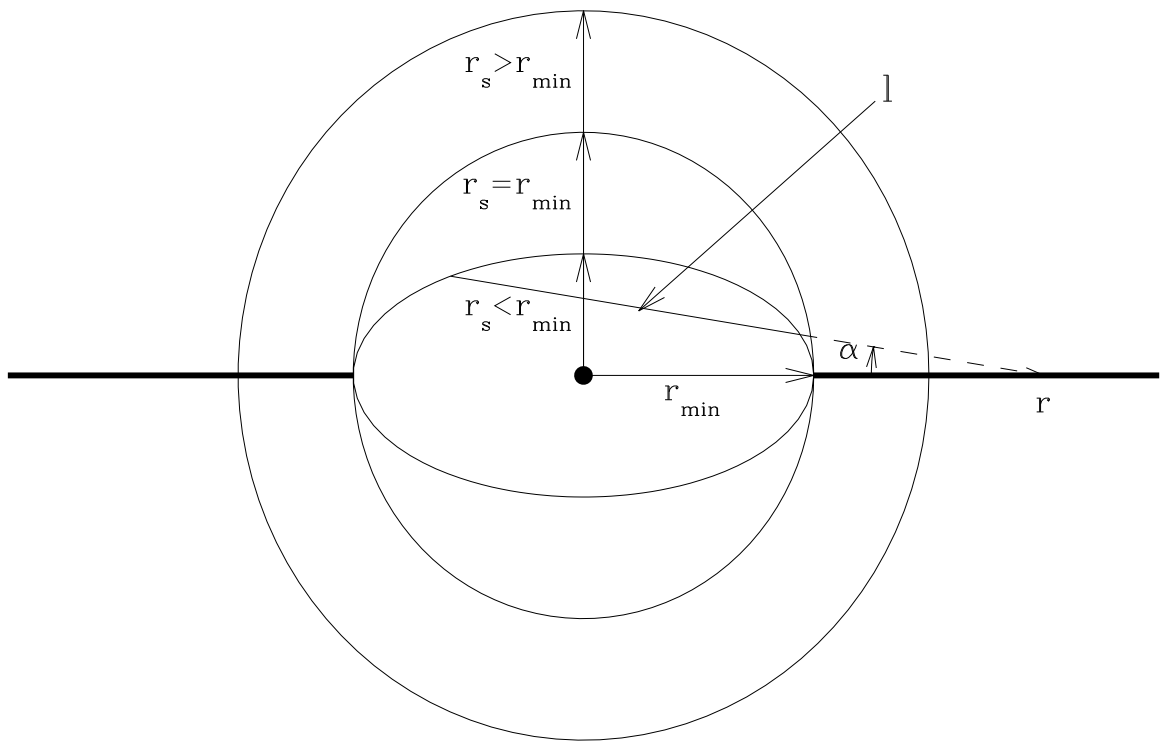}
\caption{Model geometries of the Comptonizing plasma. For $r_s \ge
\rmin$, the plasma region is a sphere, and for $r_s < \rmin$ it is an
ellipsoid with semi-major axis $\rmin$ and semi-minor axis $r_s$.  A
sample ray is shown with a path length $l$ through the ellipsoidal
Comptonizing region and making an angle $\alpha$ with respect to the
disk plane.
\label{geometry}}
\end{figure}

The inclusion of viscous dissipation has two effects on the
amplification factor.  First, as we note above, it reduces the X-ray
luminosity required to produce the observed optical emission.  Second,
it acts as a constant offset to the ZLS value of $L_s$.  Both of these
effects reduce the amplification factor for a given ratio $\rmin/r_s$
and cause the associated X-ray spectral index to be larger
(Eq.~\ref{Gamma(A)_old}).  For fits to the NGC 3516 data presented in
Paper I, the viscous dissipation needed to fit the X-ray fluxes was
large enough to require that the radius of the Comptonizing sphere had
to be smaller than the disk inner radius.  In the context of the
plasma sphere geometry, this implies a large gap between the hot
plasma and the inner edge of the thin disk.  In order to avoid such a
gap, we have modified the shape of the central hot plasma for cases
where smaller disk covering factors (i.e., smaller values of spectral
index) are required.  For $r_s < \rmin$, we take the shape of the
Comptonizing plasma to be an oblate ellipsoid with the variable $r_s$
denoting its semi-minor axis.  For $r_s \ge \rmin$, the Comptonizing
region is spherical as before.  Figure~\ref{geometry} illustrates
these two cases.

For $r_s \ge \rmin$, the radial distribution of thermal Comptonized
flux incident upon the disk is given by Eq.~\ref{Finc_sphere}.  For
$r_s < \rmin$, it is
\begin{equation}
F_{\rm inc}(r) = \frac{3L_x}{16\pi^2\rmin^2 r_s}
                 \int_0^{\alpha_\max} \sin\alpha\cos\alpha\,d\alpha
                 \int_{-\phi_\max}^{\phi_\max}
                     l\,d\phi,
\label{Finc_ellipsoid}
\end{equation}
where $\alpha$ is the angle an incoming ray makes with the disk plane,
$\phi$ is its azimuthal angle, and $l = l(r,\alpha,\phi;r_s)$ is the
path length through the Comptonizing region (see Fig.~\ref{geometry}).
Explicit formulae for $l$, $\alpha_\max$, and $\phi_\max$ are given in
the Appendix.

\begin{figure}
\plotone{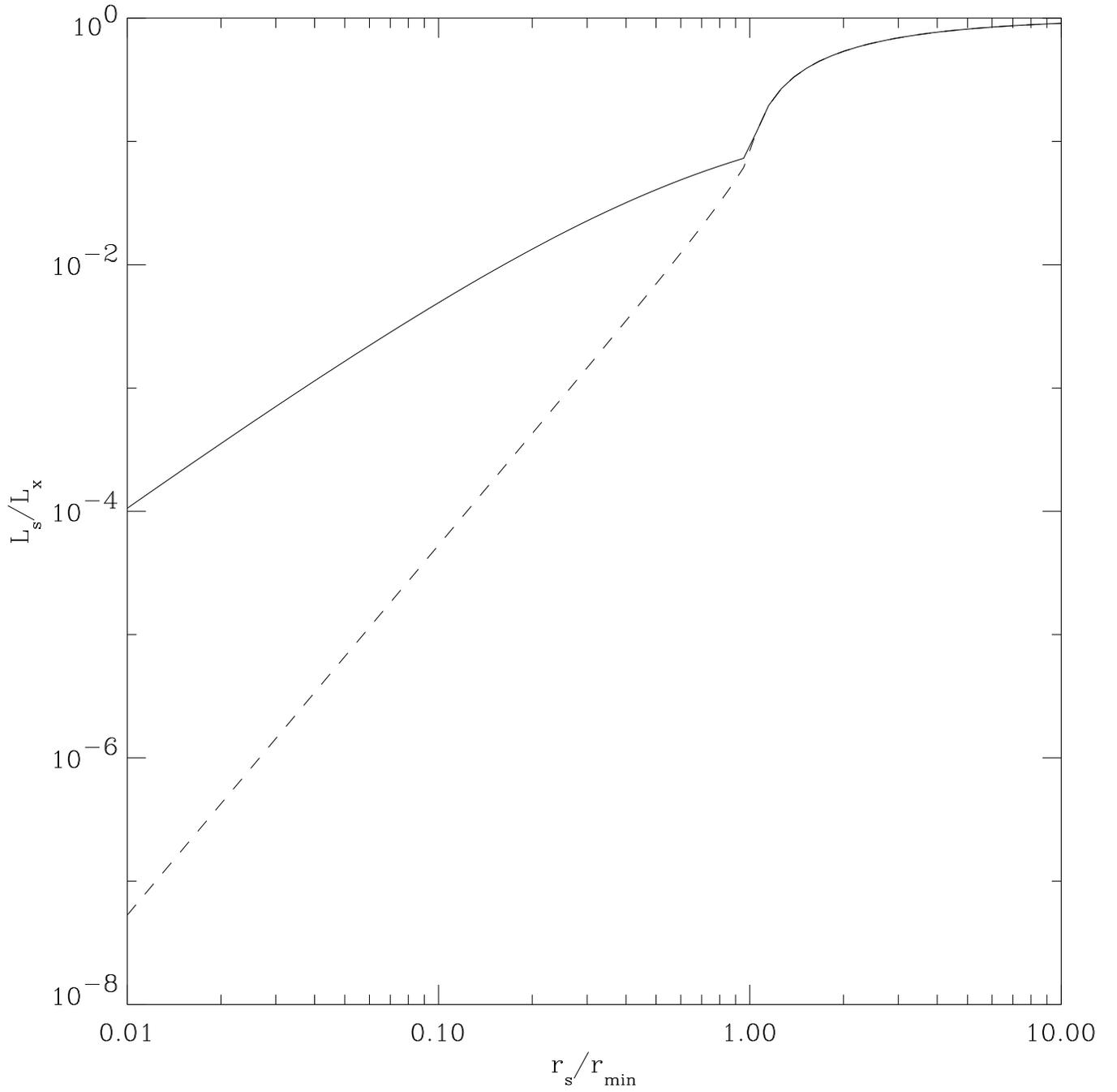}
\caption{The ratio of seed photon luminosity to thermal Comptonization
luminosity, $L_s/L_x$, versus $r_s/\rmin$ assuming zero contribution
from viscous dissipation.  The solid curve is for the hybrid
sphere/ellipsoid model, and the dashed curve is for the sphere+disk
model described by ZLS.
\label{Ls(r)}}
\end{figure}

The seed photon luminosity for the Comptonization process is the disk
flux which enters the region of hot plasma.  For our geometry, this is
given by
\begin{equation}
L_s = 4\pi\int_\rmin^\infty g(\rt)\,I(r)\,r\,dr,
\label{Ls_hybrid}
\end{equation}
where $I(r) = [(1-a)F_{\rm inc} + F_{\rm visc}]/\pi$ is the disk
surface brightness, $\rt = r/\rmin$, and we have
defined
\begin{equation}
g(\rt) = 2 \int_0^{\alpha_\max} \phi_\max(\alpha,\rt) 
               \sin\alpha\cos\alpha\,d\alpha.
\label{g(r)}
\end{equation}
In Fig.~\ref{Ls(r)}, we plot the ratio of the seed photon luminosity
to X-ray luminosity, $L_s/L_x$, as a function of $r_s/\rmin$ for the
case where the disk emission is due entirely to thermal reprocessing.
For comparison, we plot the results for the sphere+disk model
described by ZLS and the hybrid sphere/ellipsoid model.

We include an empirical color correction for the disk thermal emission
as a function of radius.  Hubeny et al.\ (2000, 2001) have modeled
disk atmospheres appropriate for Seyfert nuclei and find that the
optical continuum can be moderately well approximated by local
blackbody emission, but at UV wavelengths and higher, significant
deviations occur.  This also seems to be true for disks powered solely
by external illumination (Sincell \& Krolik 1997).  Since we are
modeling the UV emission for NGC 7469, a color correction, in principle,
should be applied.  We adopt a temperature dependent color correction of
the form
\begin{equation}
f_{\rm col}(T_\disk) = f_\infty - \frac{(f_\infty - 1) (1 +
                     \exp(-\nu_b/\Delta\nu))} { 1 +
                     \exp((\nu_p -\nu_b)/\Delta\nu)},
\end{equation}
where $\nu_p \equiv 2.82k_B T_\disk/h$ is the peak frequency of a
blackbody with temperature $T_\disk$.  This expression for $f_{\rm
col}$ goes from unity at low temperatures to $f_\infty$ at high
temperatures with a transition at $\nu_b \approx \nu_p$.  We find that
$f_\infty = 2.3$ and $\nu_b = \Delta\nu = 5\times 10^{15}$\,Hz do a
reasonable job of reproducing the model disk spectra of Hubeny et al.\
(2001).

We fit the observed optical, UV and X-ray data in the following
manner: For a given trial value of plasma temperature, $T$, the
thermal Comptonization continuum is modeled as an exponentially
cut-off power-law:
\begin{equation}
L_E = (L_0/E_c) (E/E_c)^{1-\Gamma} \exp(-E/E_c),
\end{equation} 
where we set $E_c = 2k_BT$ appropriate for plasma Thomson depths $\tau
\la 1$.  The normalization $L_0$ is given by
\begin{equation}
F_{2-10} = \frac{L_0}{4\pi d_l^2}\int_{2\,{\rm keV}/E_c}^{10\,{\rm keV}/E_c} 
           \pfrac{E}{E_c}^{1-\Gamma} \exp(-E/E_c) d(E/E_c),
\label{F210}
\end{equation}
where $F_{2-10}$ is the measured 2--10\,keV flux and $d_l$ is the
luminosity distance to the source.  The total thermal Comptonization
luminosity is then
\begin{equation}
L_x = L_0\int_{E_\min/E_c}^\infty \pfrac{E}{E_c}^{1-\Gamma} 
             \exp(-E/E_c) d(E/E_c),
\label{Lx_def}
\end{equation}
where we have set the minimum photon energy at a nominal value of
$E_\min = 0.01$\,keV.

Having determined the thermal Comptonization luminosity for a given
plasma temperature, we find the value of $r_s$ which provides the
best-fit disk spectrum:
\begin{equation}
F_\nu = \frac{4\pi\cos i}{d_l^2}\frac{h \nu^3}{c^2 f_{\rm col}^4}
     \int_\rmin^\infty \frac{r\,dr}{\exp(h\nu/f_{\rm col} k_B T_\disk) - 1},
\end{equation}
where $i$ is the observer inclination, and the disk temperature is
given by Eq.~\ref{T(r)}.  We then use the best-fit value of $r_s$ to
compute the corresponding seed photon luminosity, $L_s$, the Compton
amplification factor $A = L_x/L_s$, and the {\em predicted} value of
the X-ray photon spectral index, $\Gamma_{\rm model}$.  In contrast to
ZLS and Paper I, we use the relation
\begin{equation}
\Gamma_{\rm model} = 2.15(A - 1)^{-1/14}
\label{Gamma(A)_new}
\end{equation}
rather than Eq.~\ref{Gamma(A)_old} of Beloborodov (1999).  Malzac,
Beloborodov, \& Poutanen (2001) obtained this parameterization for
pill-box shaped coronae atop an accretion disk using a Monte Carlo
method to compute the thermal Comptonization continuum, and we have
found a nearly identical relation from our own Monte Carlo simulations
of thermal Comptonization and disk thermal reprocessing in the
sphere+disk geometry (Chiang \& Blaes 2001b). 
The best-fit value of the plasma temperature $T$ has been
ascertained when the predicted photon spectral index matches the
measured spectral index.  The effective Thomson depth, $\tau_{\rm
eff}$, of the Comptonizing plasma is estimated by inverting another
empirical relation found by Beloborodov (1999),
\begin{equation}
\Gamma = \frac{9}{4} y^{-2/9},
\end{equation}
where the Compton $y$-parameter is $y \equiv
4\theta_e(1+4\theta_e)\tau_{\rm eff}(1+\tau_{\rm eff})$ and $\theta_e
\equiv k_BT/m_ec^2$.  Although the specific values of $\tau_{\rm eff}$
that we find have no substantive role in our calculations, they may
provide, along with the plasma temperatures, useful constraints on the
properties of the Comptonizing region for the dynamical models.

The other important model parameters are the disk inner truncation
radius, $\rmin$, the central black hole mass, $M$, and the mass
accretion rate, $\Mdot$.  We assume that these quantities are constant
throughout the monitoring campaigns for each object.  The black hole
masses can certainly be taken to be constant, and $\rmin$ and $\Mdot$
will likely only change appreciably on viscous time scales which are
on the order of ${\cal O}(10^1)$ years for typical Seyfert 1
parameters.  For NGC 7469, we have chosen a black hole mass of $M_7
\equiv (M/10^7\,$\Msun) $ = 1$.  This is consistent with the upper
range of the mass estimates by Collier et al.\ (1998) using broad
emission line reverberation mapping.  As we discussed in Paper I, the
disk inner truncation radius will be largely constrained by the shape
and magnitude of the optical/UV continuum.  For NGC 7469, the optical
and UV fluxes show evidence for the roll-over associated with the peak
of the black body spectrum at this inner radius.  Since the
temperature distribution of the disk will nearly follow a $T_\disk
\propto r^{-3/4}$ radial dependence whether the local disk flux is due
to viscous dissipation or thermal reprocessing from an extended
central X-ray source (see Paper I), we can use the optical and UV
colors and magnitudes for NGC 7469 to estimate values for the disk
inner radius $\rmin$ and the disk temperature $T_{\disk,\max}$ at that
radius.  Insofar as the radial temperature distribution obtained from
our more detailed thermal reprocessing plus viscous dissipation
calculation does obey a $r^{-3/4}$ dependence, the values we find for
$\rmin$ and $T_{\disk,\max}$ will depend {\em only} on the optical and
UV fluxes.  Performing these fits over the 30 epochs of optical/UV
data for NGC 7469, the average values we obtain are $\rmin = 3.0 \pm
0.3 \times 10^{14}$~cm and $T_{\disk,\max} = 3.5 \pm 0.2\times
10^4$~K.  As we showed in Paper I, disk spectra computed for values of
$\rmin$ which differ significantly from the above will have very
different optical/UV colors if normalized to the longest wavelength
optical flux.  As {\em a posteriori} confirmation of setting $\rmin =
3 \times 10^{14}$~cm for NGC 7469, we note the quality of the fits to
the optical/UV data shown in Fig.~\ref{7469_SEDs} and that the
temperature distributions for those calculations do indeed have close
to a $r^{-3/4}$ dependence.  Now, since $\rmin \ga 200\,r_g$, the
amount of disk emission due to viscous dissipation is insensitive to
the term containing $r_I$ in Eq.~\ref{Fvisc(r)} and depends almost
entirely on the product $M\Mdot$.  Therefore, the full range of black
hole masses, $M \sim 10^6$--$10^7$\,\Msun, found by Collier et al.\
can be accommodated by adjusting $\Mdot$ accordingly.

Simultaneous {\em ASCA} measurements of the broad Fe K$\alpha$ line
emission in NGC 3516 indicate that much of this emission originates
from disk radii $\la 6\,r_g$, very close to the central black hole
(Nandra et al.\ 1999).  Therefore we adopt a disk inner radius of
$\rmin = 6\,r_g$, so that the choice of black hole mass also sets the
value of $\rmin$.  Unlike NGC 7469, the optical and UV continuum
fluxes we have for NGC 3516 do not show a high frequency roll-over,
and we can infer only an upper limit for the disk truncation radius
from these data.  After de-reddening the optical and UV fluxes, a disk
inner truncation radius of $\rmin \la 1 \times 10^{13}$\,cm seems to
be required, yielding a black hole mass of $M \la
10^7$\,\Msun.\footnote{In Paper I, the effects of reddening were not
considered, and a truncation radius of $\rmin \simeq 3\times
10^{13}$\,cm was sufficient.}  However, a mass this small yields
somewhat large values for the temperature of the Comptonizing plasma
during some epochs (see Fig.~\ref{3516_lcs}), so we have computed
models for several black hole masses $M_7 = 1$, 2, and 3.

Given values for $\rmin$ and $M$, several considerations limit the
plausible range of accretion rates.  First, since they are assumed to
be constant, they must be sufficiently small to allow the observed
optical and UV variability to be due to disk thermal reprocessing.
Second, since the amount of disk flux and the ratio $r_s/\rmin$
determine the X-ray luminosity through the observed photon index and
amplification factor scaling relation, larger accretion rates
correspond to smaller thermal Comptonization roll-over energies, $E_c
= 2k_BT$.  The values of $E_c$ cannot be too small, otherwise the
roll-over would be evident in the observed X-ray spectra.  Therefore,
somewhat arbitrarily, we selected accretion rates which ensured that
$k_B T \ga 50$~keV in all epochs.  As it turns out, this condition is
more stringent than the constraint that the varying part of the
optical/UV flux need be due to thermal reprocessing.  On the other
hand, the accretion rate must be sufficiently large so that energy
balance is satisfied such that $L_{\rm untr} \ga L_x + L_\disk$.
Furthermore, larger accretion rates are preferred since they imply
values of $k_B T$ which are not too large.  This is essential since we
do not consider the effects of pair balance or emission from
non-thermal particle distributions, and thus this model tacitly
assumes $k_B T \la m_e c^2$.  The various parameters which are held
constant are summarized in Tables~\ref{3516_parameters}
and~\ref{7469_parameters}.

\section{Results}
\subsection{NGC 3516}

\begin{figure}
\plotone{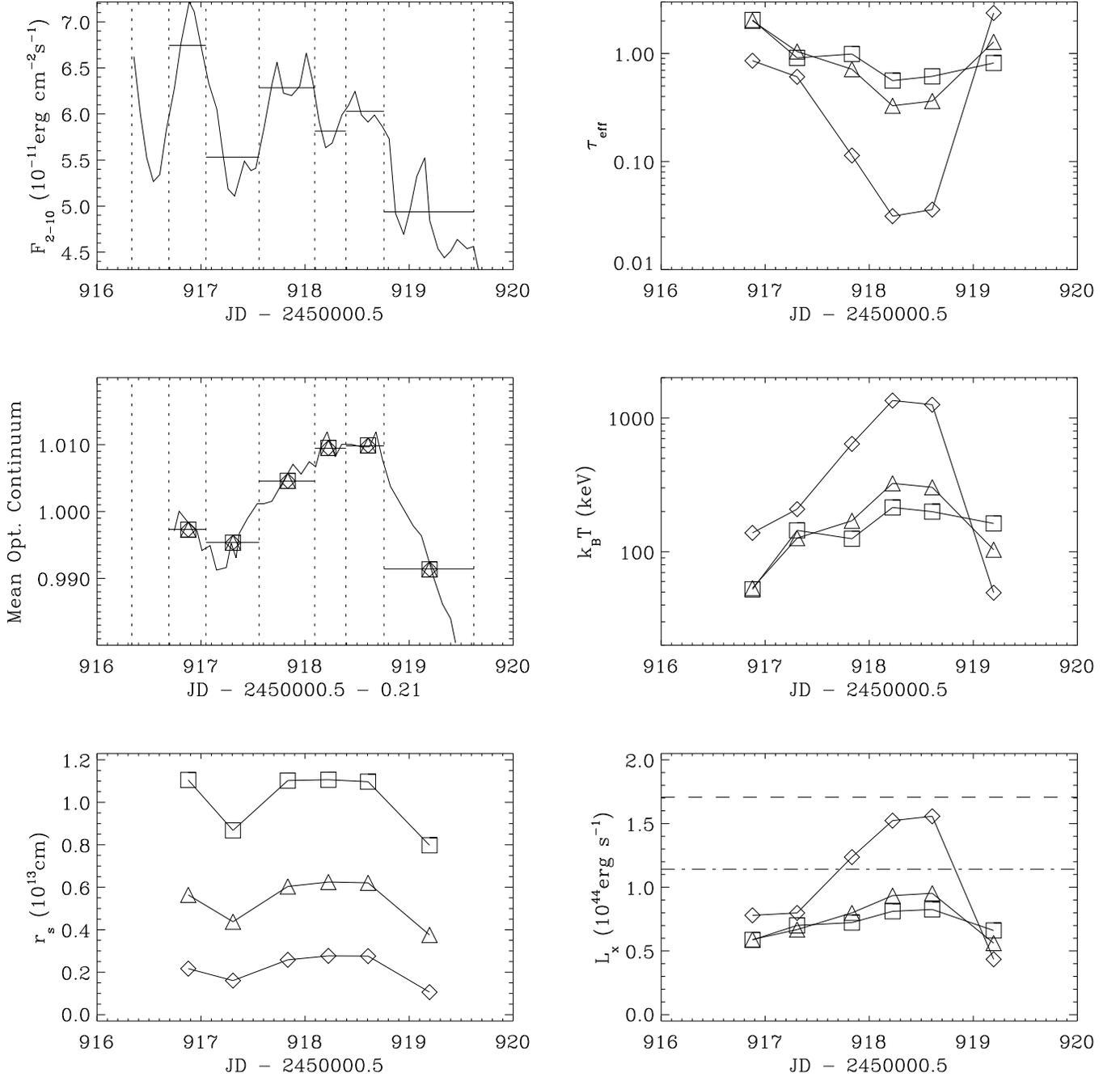}
\caption{Model fluxes and fitted parameters for NGC 3516. The upper
left plot shows the 2--10 keV X-ray light curve (solid curve). The
dotted vertical lines are the boundaries of the epochs used to
determine the X-ray spectral indices, and the horizontal line segments
are the mean fluxes that were used to compute $L_x$ (see
Eqs.~\ref{F210} and~\ref{Lx_def}).  The middle left plot shows the
optical continuum light curve (solid curve) averaged over the fluxes
at 3590\AA, 4235\AA, and 5510\AA\ and scaled to unity.  Also shown are
the mean values for each epoch (line segments), and the fitted values
for $M_7 = 1$ (diamonds), 2 (triangles), and 3 (squares).  The lower
right plot shows the thermal Comptonization luminosities compared to
the available accretion power for $M_7 = 2$ (dashed line) and $M_7 =
3$ (dot-dashed line).  The value for $M_7 = 1$ is $L_{\rm untr} -
L_\disk = 3.36 \times 10^{44}$\,erg~s$^{-1}$ (see
Table~\ref{3516_parameters}).
\label{3516_lcs}}
\end{figure}

The model parameters and resulting light curves for NGC 3516 are shown
in Fig.~\ref{3516_lcs}.  For the $M_7 = 2$ and $M_7 = 3$ cases, the
fitted plasma temperatures and optical depths are well within the
range of values which have been typically measured or assumed for type
1 Seyfert galaxies (Zdziarski, Poutanen, \& Johnson 2000).  However,
for the $M_7 = 1$ case, the plasma temperature is somewhat high in two
epochs, $k_B T \sim 1$\,keV.  In all three cases, the size parameter
of the Comptonizing region, $r_s$, is about a factor of 3--4 smaller
than the disk inner truncation radius so that the shape of the region
containing the hot plasma is relatively flat.  By comparison, the
range of spherical radii we found for the fits presented in Paper I
were only somewhat smaller than the truncation radius, $r_s \sim
0.7\,\rmin$.  The reason for this less pronounced difference can be
seen in Fig.~\ref{Ls(r)} in which the $L_s/L_x$ curve for the ZLS
calculation declines more rapidly as $r_s/\rmin$ decreases for
$r_s/\rmin < 1$ than it does for the hybrid sphere/ellipsoid model.
In the lower right panel of Fig.~\ref{3516_lcs}, we see that the
thermal Comptonization luminosities, $L_x$, are consistently less than
the available accretion power, $L_{\rm untr} - L_\disk$ (see
Table~\ref{3516_parameters}).  For the black hole masses and
accretion rates we consider, the thermal Comptonization luminosity
constitutes a substantial fraction of $L_{\rm untr} - L_\disk$,
implying fairly efficient accretion, comparable to that for an
untruncated thin disk in a Kerr metric.

\begin{figure}
\plotone{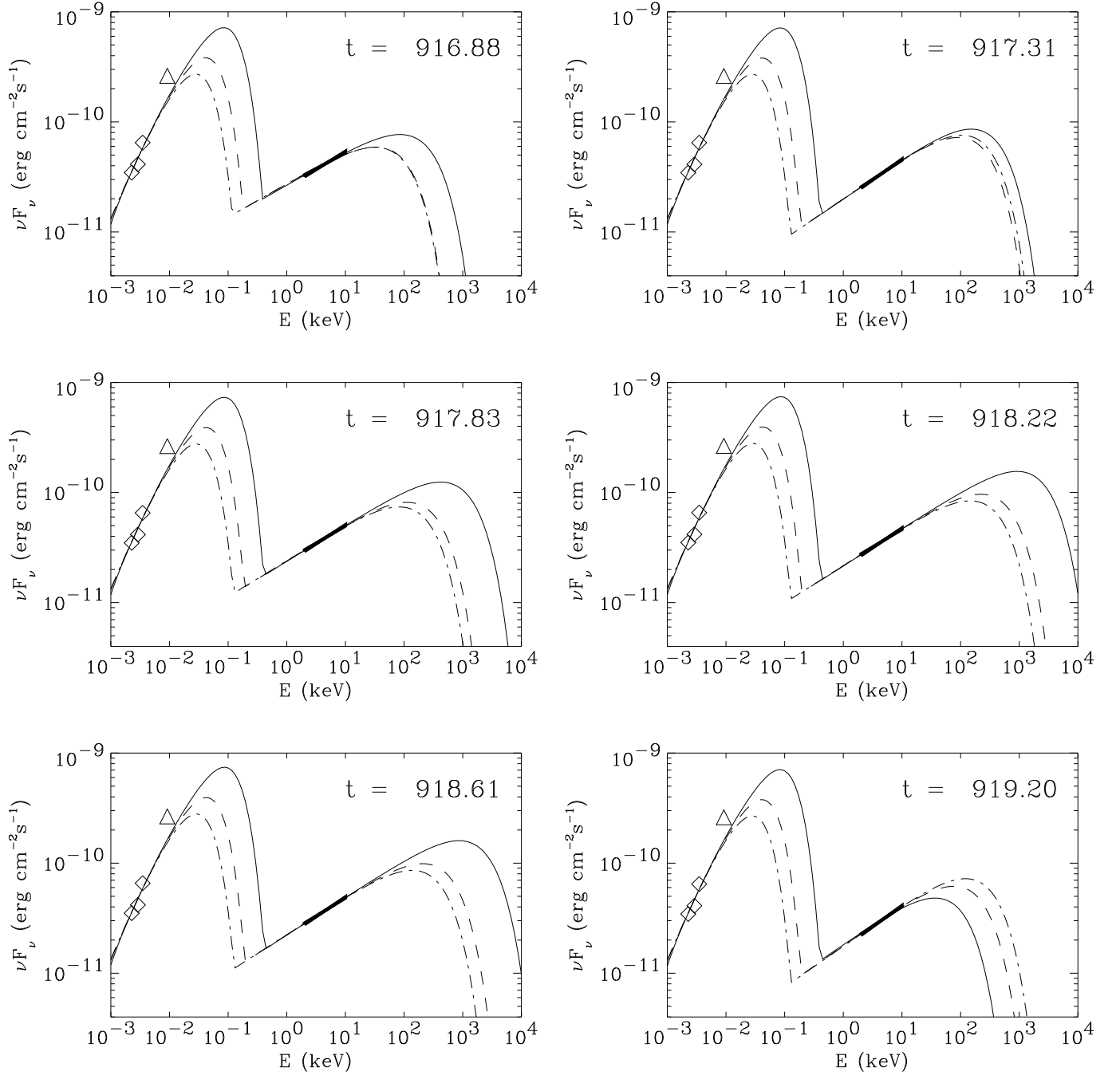}
\caption{Model spectral energy distributions for NGC 3516 compared to
the observed spectra. The SEDs are show for $M_7 = 1$ (solid curve), 2
(dashed), 3 (dot-dashed).  The diamonds are the de-reddened optical
fluxes from Edelson et al.\ (2000) which have been used in the fits to
the disk thermal spectra.  The triangle in each plot is the
de-reddened non-simultaneous flux at 1360\AA.  The thick line segments
are the 2--10 keV power-law continua which we have fit to the archival
{\em RXTE} data.
\label{3516_SEDs}}
\end{figure}

The spectral energy distributions for these fits are shown in
Fig.~\ref{3516_SEDs}.  The optical and X-ray spectra are reasonably
well described by the model SEDs, but the continuum flux at 1360\AA\
is problematic.  When the reddening correction is applied, this datum
sits above a $\nu F_\nu \propto \nu^{4/3}$ extrapolation from the
optical data.  Even if we made the model disk spectra bluer by, for
example, decreasing the inner truncation radius below $\rmin = 8.9
\times 10^{12}$\,cm, the 1360\AA\ flux would still exceed the model
value by $\sim 20$\%.  Edelson et al.\ (2000) estimate systematic
changes in the UV flux of 1.5\% due to thermal variations, but this is
too small to account for the potential discrepancy.  Furthermore,
given the lack of a strong X-ray flare either during or preceding the
UV measurements (see Fig.~1, Edelson et al.\ 2000) and the $\pm 1.5$\%
relative variations in the optical light curves, which directly
followed them, it is unlikely that there was a sufficiently strong
flare in the disk emission to account for such a large UV continuum
flux.  However, the 1360\AA\ continuum may be contaminated by the
broad wings of the Si\,{\sc iv} 1397\AA\ and C\,{\sc ii} 1335\AA\
emission lines.  It would be extremely useful for a more complete
analysis of these data to be performed that is similar to the spectral
fitting by Kriss et al.\ (2000) for the {\em HST/Faint Object
Spectrograph} and {\em IUE} spectra of NGC 7469 in order to determine
the separate continuum and emission line fluxes as precisely as
possible.

\subsection{NGC 7469}
\begin{figure}
\plotone{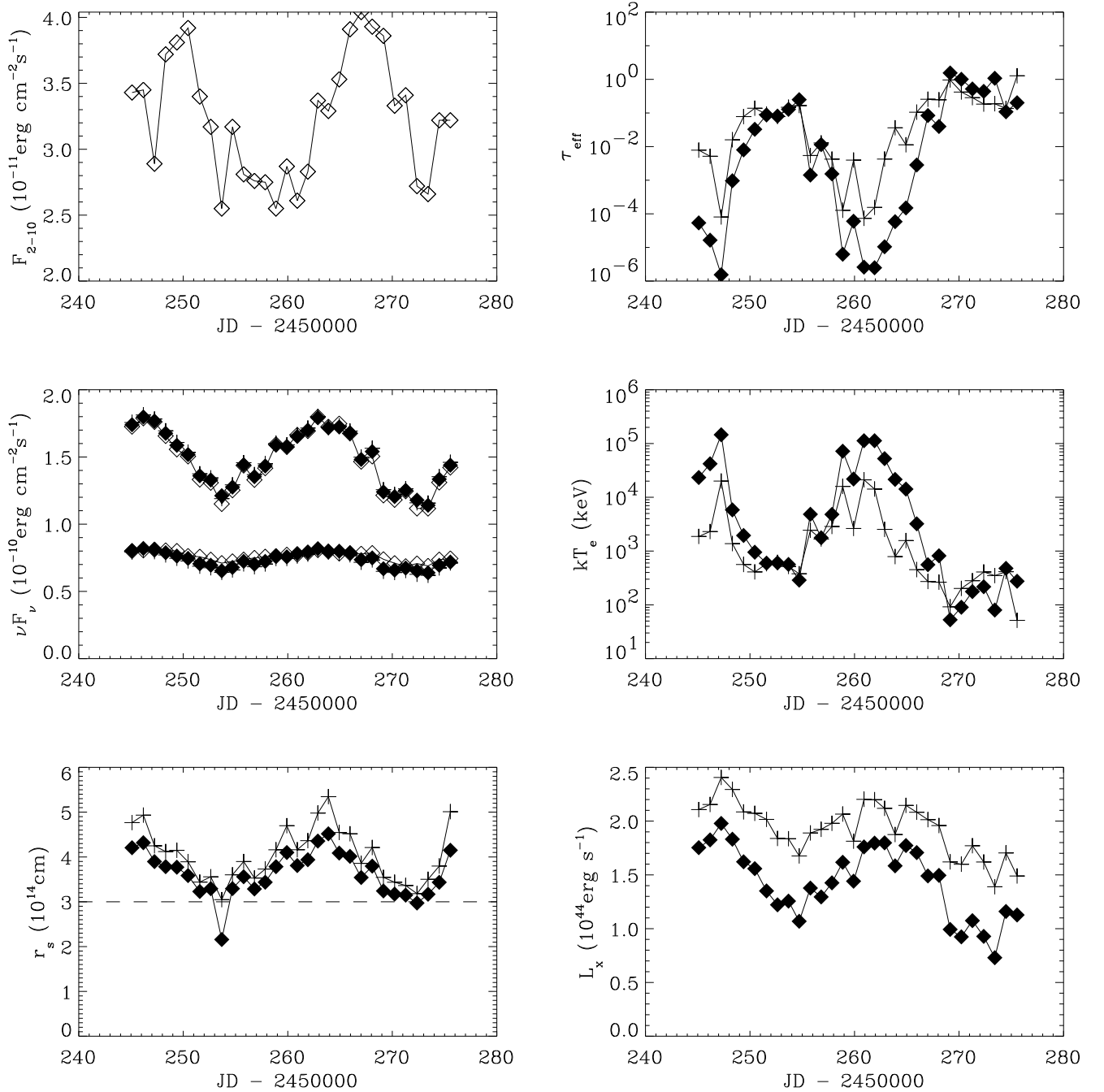}
\caption{Model light curves and fitted parameters for NGC 7469 using
anisotropy parameter values $\xi = 1$ (filled diamonds), corresponding
to isotropic thermal Compton emission, and $\xi = 1.5$ (plus-signs);
see Eq.~\ref{Lx_app}. In the middle left plot, the measured fluxes are
the open diamonds joined by line segments.  The upper set of curves
are the 1315\AA\ fluxes, and the lower set of curves are the 4865\AA\
fluxes.  The dashed line in the lower left panel is the disk inner
truncation radius, $\rmin = 3 \times 10^{14}$\,cm.
\label{7469_lcs}}
\end{figure}

The corresponding fits for NGC 7469 shown in Fig.~\ref{7469_lcs}
(solid diamonds) present a different picture for the inner regions of
this object and seem to require additional elements in the model.
First of all, the generally softer X-ray spectra imply values of $r_s$
which are close to the disk inner radius so that the central plasma
region is spherical, save for one epoch when the X-ray spectrum is its
hardest.  Of course, the specific values we fit for $r_s$ also depend
on our choice of $\Mdot$.  Second, the variations in the observed
optical fluxes are noticeably smaller than that of the model.  Part of
the excess variability in the model optical light curves is due to the
fact that light travel time effects are not included in these
calculations.  The measured time lags of $\sim 1$ day imply that the
model curves should be smeared out on at least this time scale.
However, even if such light travel time effects are included, there
will still be excess variability at the $\sim 10$ day time scale.
Nonetheless, given the simplicity of this model, it is encouraging
that the shape and magnitude of the UV light curve can be so well
reproduced while the relative magnitude of the optical fluxes are
approximately matched and the optical variability is not greatly
over-predicted.

Unfortunately, the relatively large variations in the UV flux seem to
require much larger changes in the X-ray luminosity than are found by
simply extrapolating the 2--10 keV flux over even as broad a range as
0.1 to 100 keV (see Nandra et al.\ 2000).  Given the measured
2--10~keV fluxes and photon indices, the size of the $L_x$ variations
which are required by thermal reprocessing can only be accommodated by
setting the thermal roll-over energy to unrealistically large values
--- in several epochs, we find $k_BT \ga 10^5$~keV.  Clearly, such
large temperatures are not consistent with the underlying assumptions
of these calculations.  Not only would pair-production and non-thermal
emission likely be present at these temperatures (Coppi 1999), the
shape of the thermal Comptonization continuum at 2--10 keV would not
resemble a power-law as individual scattering orders would become
apparent in these spectra for temperatures $\ga 10^3$~keV (e.g., Stern
et al.\ 1995).

One way of narrowing the required temperature range is by giving the
observed X-ray continuum more ``leverage'' in producing a greater
amount of variability in the thermal reprocessed disk emission.  This
can be accomplished by allowing for anisotropy in the thermal Compton
emission so that the X-ray intensity that we see directly is less than
the X-ray intensity incident upon the accretion disk.  Such anisotropy
may arise in several different ways.  The strong gravitational field
of the central black hole will cause radiation produced near it to
follow curved geodesics which may intercept the disk rather than
follow straight-line trajectories to infinity (e.g., Cunningham 1975).
Bulk relativistic motion of the Comptonizing plasma will beam the
radiation along the direction of motion (Beloborodov 1999).  However,
in this case, the bulk motion would have to be directed towards the
disk instead of away from it as would occur in a mildly relativistic
jet.  Finally, since the region of Comptonizing plasma is not being
illuminated uniformly, but rather by the seed photons from the disk
encircling its equator, there is a strong preference for the thermal
Compton photons of the first scattering order to be directed back
towards the disk.

Ideally, we would have full descriptions of the geodesics followed by
the thermal Compton radiation in order to account for the
aforementioned effects properly, but the use of a simple
parameterization will serve to illustrate how large a difference some
mild anisotropy can make.  Therefore, rather than modeling the various
effects in detail, we instead introduce a parameter, $\xi$, which
describes the anisotropy so that the total luminosity of the
Comptonizing plasma and the luminosity used to compute the thermal
reprocessed emission is given by
\begin{equation}
L_x = \xi L_{x,{\rm app}}.
\label{Lx_app}
\end{equation}
Here $L_{x,{\rm app}}$ is the apparent luminosity inferred from the
assumed thermal roll-over, $E_c = 2k_BT$, the measured spectral index,
and the measured 2--10~keV flux, i.e., it is the $L_x$ given by
Eqs.~\ref{F210} and~\ref{Lx_def}.  Setting $\xi = 1.5$, we have
recomputed the spectral fits to the NGC 7469 data and plotted the
results in Fig.~\ref{7469_lcs} (plus-signs).  As expected, a narrower
range of plasma temperatures are found which reproduce the optical/UV
variability, but the plasma temperatures nonetheless reach values
$k_BT \sim 10^4$~keV, which are still too large to be self-consistent.

\begin{figure}
\plotone{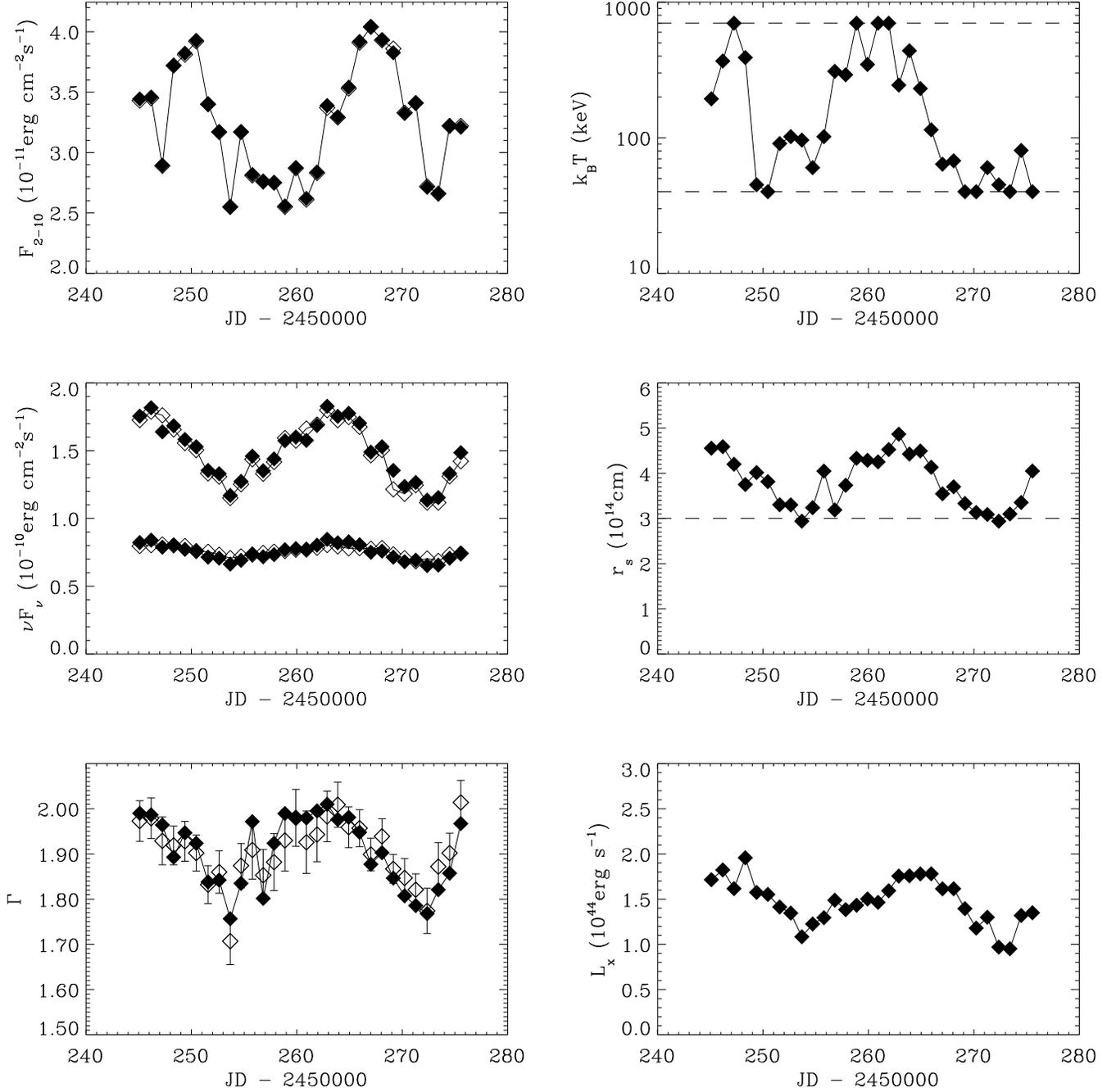}
\caption{Model light curves and fitted parameters for NGC 7469 using
$\xi = 1.5$ and allowing the X-ray spectral indices for each epoch to
vary within the measured 1-$\sigma$ bounds. The open diamonds are the
measured or best-fit values determined from the data while the filled
diamonds are the model values or parameters found from our spectral
fits.
\label{7469_lcs_fudge}}
\end{figure}

\begin{figure}
\plotone{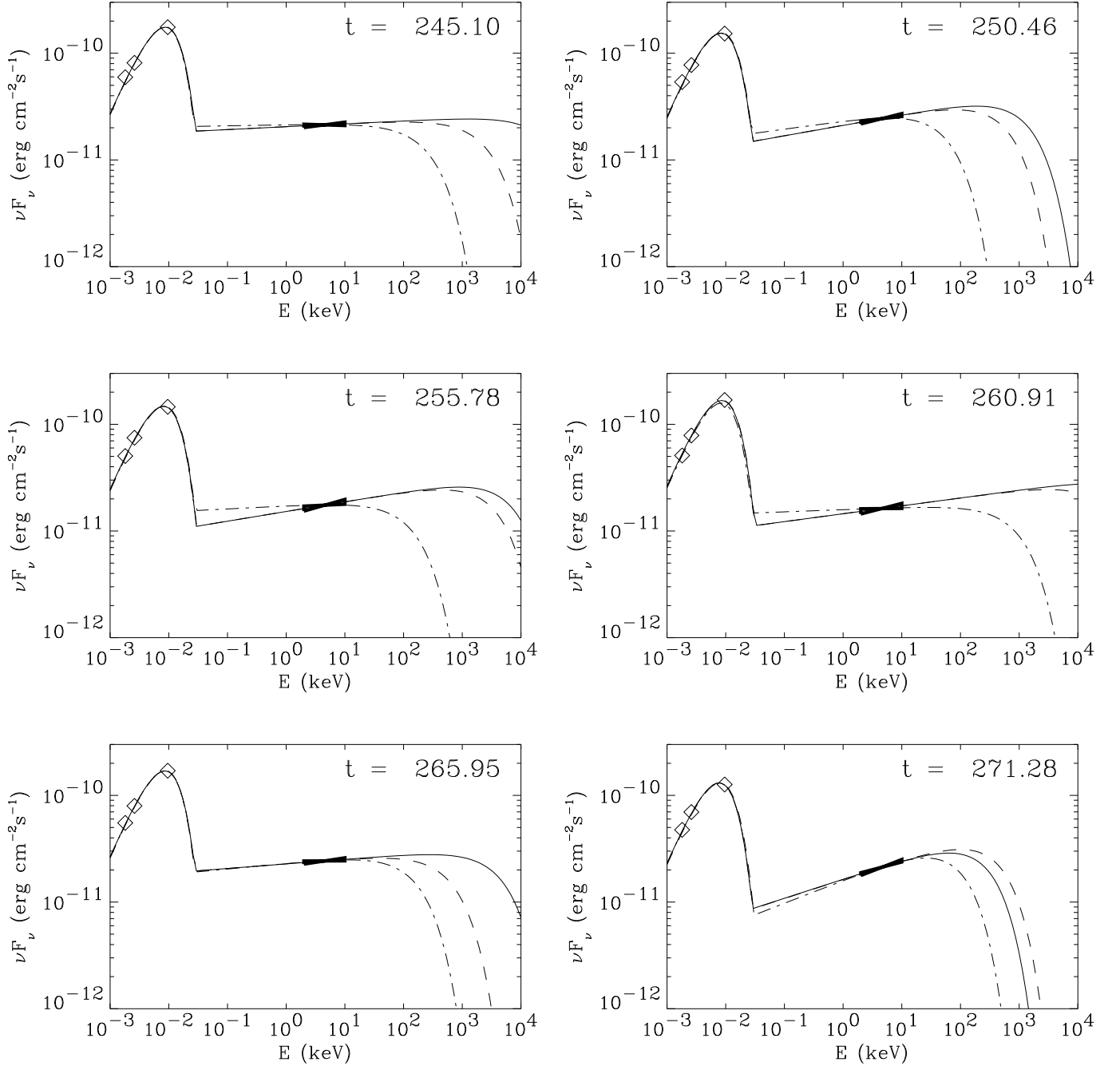}
\caption{A subset of the model spectral energy distributions for NGC
7469.  The solid and dashed curves correspond to $\xi = 1$ and $\xi =
1.5$, respectively, using fixed values for the X-ray spectral indices,
$\Gamma$. The dot-dashed curves are the SEDs found by allowing the
$\Gamma$-values to vary within the 1-$\sigma$ bounds and using $\xi =
1.5$.  The open diamonds are the starlight subtracted, de-reddened
optical and UV fluxes from 1996 monitoring campaign.  The thick line
segments are the 2--10 keV power-law continua from the fit parameters
given in Nandra et al.\ (2000).
\label{7469_SEDs}}
\end{figure}

We now ease some of the restrictions on the parameters of the X-ray
spectrum.  For all previous fits, we have used the best-fit values of
the photon indices obtained from the X-ray spectral fitting.  Since
many of the spectral indices for the NGC 7469 observations are near
$\Gamma \approx 2$ (see Fig.~\ref{7469_lcs_fudge}), a given change in
the thermal roll-over energy $E_c$ has less impact on the overall
X-ray luminosity than it would for a harder spectrum with, say,
$\Gamma \approx 1.6$.  Therefore, in order to increase the accessible
range of the X-ray luminosities, we allow the X-ray spectral index to
vary within the 1-sigma bounds of the best-fit values, and we allow
for small variations in the 2--10 keV flux used to compute $L_x$
(Eq.~\ref{F210}).  Furthermore, we restrict the plasma temperature to
be $40$~keV $< k_BT < 700$~keV; and we set the anisotropy parameter to
be $\xi = 1.5$, as before.  With these additional conditions, we
obtain the spectral fits shown in Fig.~\ref{7469_lcs_fudge}.  For some
epochs, the X-ray spectral indices attain values which are right at
the 1-sigma limits, and small deviations from the observed X-ray
fluxes and somewhat larger deviations in UV fluxes are now present,
but the overall fits of the spectral energy distributions are largely
satisfactory while maintaining a consistent range of plasma
temperatures.  We plot in Fig.~\ref{7469_SEDs}, the SEDs for six
exemplary epochs of the NGC 7469 data for the three cases we have
considered.  The details of these three fitting schemes are given in
Table~\ref{7469_parameters}.

\section{Discussion and Conclusions}
Given the parameters from the above spectral fits, NGC 3516 and NGC
7469 provide contrasting cases for the characteristics of the
accretion flow in the inner regions of Seyfert galaxies.  NGC 3516 has
a high accretion efficiency ($\ga 30$\%), a small disk inner
truncation radius, and a flattened, disk-like distribution of
Comptonizing plasma.  On the other hand, NGC 7469 has a substantially
lower accretion efficiency ($\sim 2$--3\%), a large truncation radius,
and a vertically extended hot plasma region.  As we have noted above,
the small truncation radius for NGC 3516 is consistent with the
relativistically broadened iron K$\alpha$ line measured by {\em ASCA}
over the same observing period.  Although similar {\em ASCA}
observations were not taken during the 1996 campaign for NGC 7469,
Nandra et al.\ (2000) found iron K$\alpha$ equivalent widths of $\sim
150$\,eV from the {\em RXTE/PCA} data.  This value agrees with the
equivalent width measured during the November 1993 {\em ASCA}
observations by Guainazzi et al.\ (1994) who found the iron K$\alpha$
emission line to be very narrow.  If the Fe~K$\alpha$ line is composed
of emission from a disk, then Guainazzi et al.\ concluded that it must
originate at least several tens of Schwarzschild radii from the
central black hole, consistent with our estimate for the disk inner
truncation radius of $\rmin \simeq 200\,r_g$.

As we discussed in the Introduction, a similar analysis was performed
by Poutanen et al.\ (1997) for the X-ray binary Cygnus X-1.  With
respect to the semi-analytic model they used, the main difference
between that work and the present is in the empirical relation
employed by those authors to characterize the thermal Comptonization
spectrum in terms of the model parameters.  In particular, rather than
Eq.~\ref{Gamma(A)_new}, Poutanen et al.\ used the scaling relation
\begin{equation}
\Gamma = \pfrac{20}{3A}^{1/4} + 1
\label{Gamma(A)_PKR}
\end{equation}
which was found from Comptonization modeling by Pietrini \& Krolik
(1995).\footnote{Eq.~\ref{Gamma(A)_PKR} differs numerically from the
expression used by ZLS, Eq.~\ref{Gamma(A)_old}, by only $\la 3$\% for
photon indices ranging from $\Gamma = 1.4$ to 2.3.}  On the
observational side, since Cygnus X-1 is significantly brighter in the
hard X-rays than any Seyfert galaxy, both the roll-over energy of the
thermal Comptonization continuum and the amplitude of the Compton
reflection component for Cygnus X-1 could be measured directly.  These
data provided Poutanen et al.\ with estimates of the plasma
temperature as well as the ratio $r_s/\rmin$.  In addition, the
thermal disk spectrum, which was observable in the soft X-rays, gave
them constraints on the accretion disk flux and the inner disk radius.
Armed with this information, Poutanen et al.\ were able to apply the
scaling relation and the assumed sphere+disk geometry to determine the
mass accretion rate directly in both the hard and soft spectral states
rather than to have to assume values for the accretion rates based on
``reasonable'' ranges for parameters such as the plasma temperature,
as we have done here.

\begin{figure}
\plotone{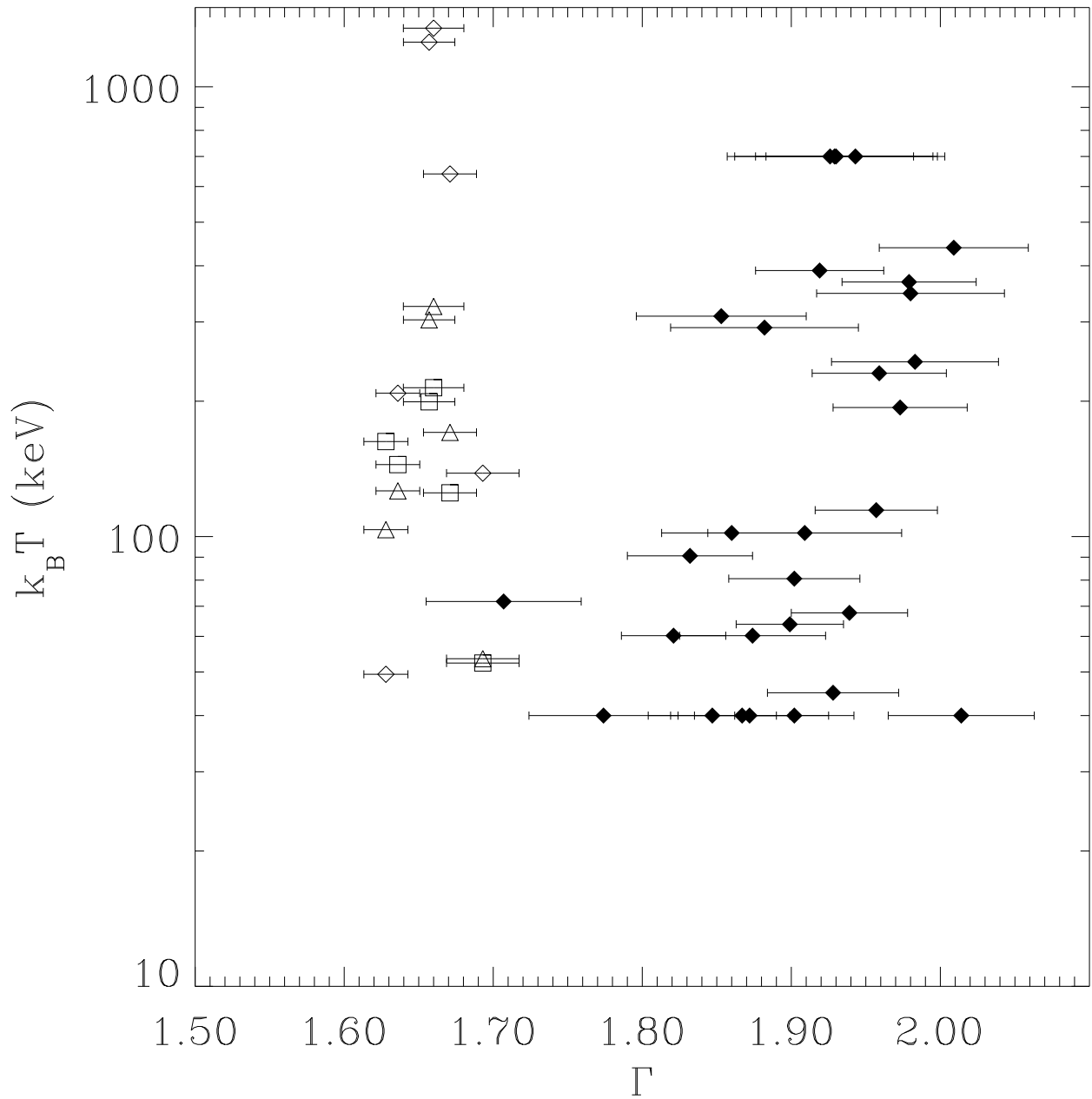}
\caption{Model plasma temperatures versus measured X-ray spectral
indices.  The open symbols are the NGC 3516 best-fit values for $M_7 =
1$ (diamonds), 2 (triangles), and 3 (squares), and the filled diamonds
are the values obtained for NGC 7469 using $\xi = 1.5$ and allowing
$\Gamma$ to vary.
\label{kT_vs_Gamma}}
\end{figure}

As the analysis of the Cygnus X-1 data demonstrates, more stringent
tests of this model in its application to Seyfert galaxies will be
available when the hard X-ray/soft gamma-ray spectra from these
objects can be reliably measured on sufficiently short time scales to
detect the changes in the spectral roll-over implied by the
temperature curves in Figs.~\ref{3516_lcs}, \ref{7469_lcs},
and~\ref{7469_lcs_fudge}.  This will require a hard X-ray telescope
with a large effective area at these energies such as that proposed
for {\em ASTRO-E2}.  Nonetheless, relevant constraints for the mean
properties of the Comptonizing plasma may be ascertained from existing
longer time scale observations by the {\em Beppo-SAX} satellite and
the {\em Oriented Scintillation Spectrometer Experiment (OSSE)} aboard
the {\em Compton Observatory} and from forthcoming observations which
will be available with {\em INTEGRAL}.  For example, spectral fits of
data from {\em OSSE} observations of NGC 4151 (Johnson et al.\ 1997)
and from {\em Beppo-SAX} observations of NGC 5548 (Nicastro et al.\
2000) suggest that as the X-ray power-law spectrum softens, the
cut-off energy (or equivalently, the plasma temperature) increases,
contrary to naive expectations from thermal Comptonization models.
However, since the Thomson depth of the plasma may also change
simultaneously, a sufficiently large decrease may more than compensate
for any spectral hardening which a temperature increase would
otherwise imply.  In Fig.~\ref{kT_vs_Gamma}, we plot the plasma
temperatures versus the measured X-ray spectral indices for the
``optimal'' case fits of the NGC 7469 data (i.e., $\xi = 1.5$ and
$\Gamma$ allowed to vary) and for the three cases we considered for
NGC 3516.  It is clear that a strict relationship between $k_BT$ and
$\Gamma$ is not required. Since changes in $k_B T$ will affect both
$L_x$ and the thermal reprocessing contribution to the disk emission,
simultaneous optical and UV monitoring will be necessary in order for
longer term measurements of the hard X-ray spectra of Seyfert 1s to
constrain thermal Comptonization models in which thermal reprocessing
plays an important role.

Greater spectral coverage in the hard X-ray band will also provide
improved constraints on the apparent correlation between the thermal
Comptonization spectral index and the relative strength of the Compton
reflection component.  However, the predictions of the ZLS analysis
are altered since the inclusion of disk emission due to viscous
dissipation implies quite different estimates for the seed photon
luminosities.  The X-ray spectral indices will no longer be solely a
function of the geometry, and simultaneous optical and UV monitoring
along with broad band X-ray spectroscopy will again be necessary to
determine the strength of the Compton reflection component implied by
this model.

\begin{figure}
\epsscale{0.6}
\plotone{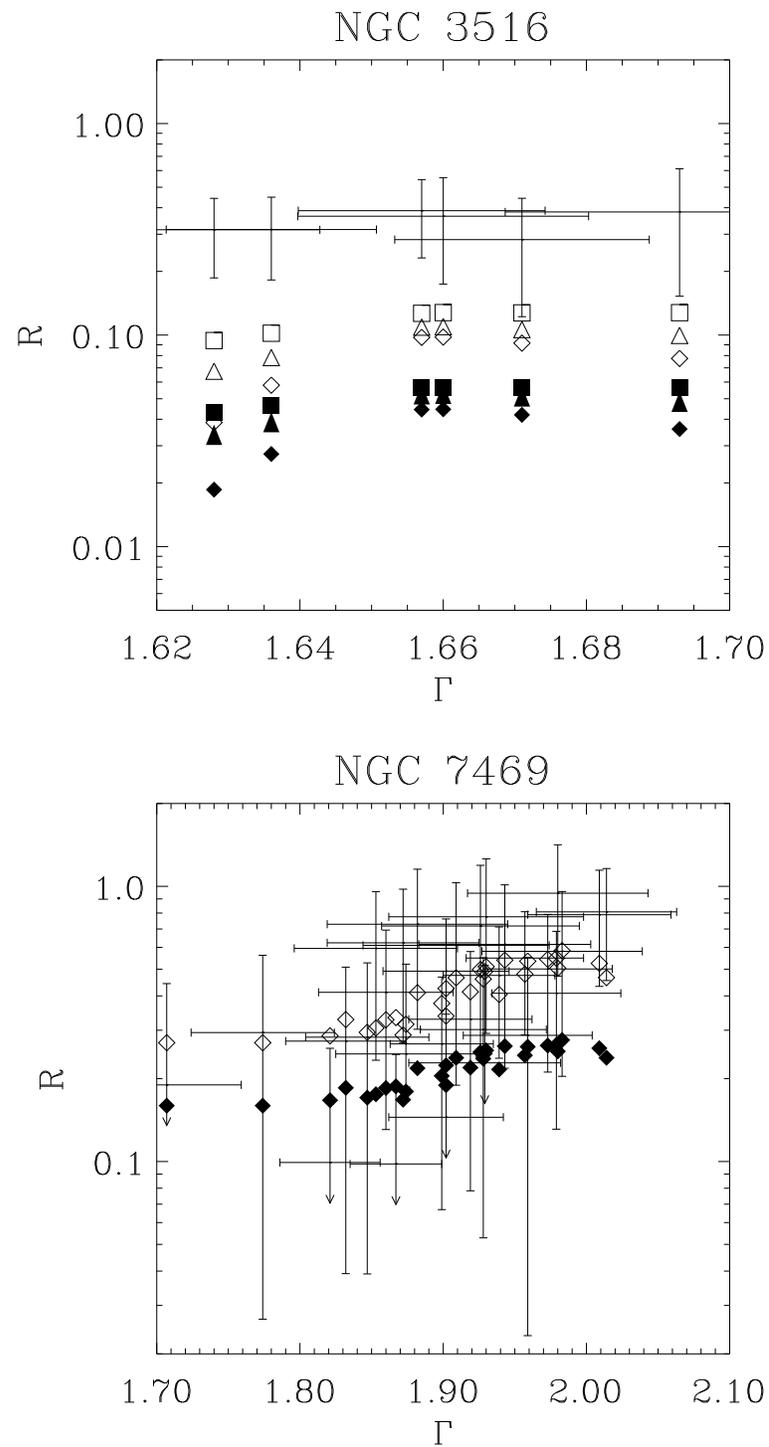}
\caption{Compton reflection fraction versus X-ray spectral index for
NGC 3516 (upper figure) and NGC 7469 (lower). The filled symbols are
for $R_{i=0} = L_d/2L_{x,{\rm app}}$ which is the implied reflection
fraction for a face-on geometry, while the open symbols are for
$R_{\rm ZLS} = L_d/(L_x - L_d)$ which is the {\em mean} reflection
fraction as defined by ZLS.  The reflection fractions and spectral
indices obtained from fitting the {\sc pexrav} model to the {\em
RXTE/PCA} data are shown with the error bars.
\label{Gamma_vs_R}}
\end{figure}

In fact, for the present data, we do find significant differences
between the predicted reflection fractions and the values obtained by
fitting the {\sc pexrav} model to the {\em RXTE/PCA} data.  In
addition, the reflection fraction given by this model for a specific
viewing angle can differ substantially from the {\em average} Compton
reflection strength as defined by ZLS.  Following ZLS, the total
thermal Comptonization luminosity intercepted by the disk is
\begin{equation}
L_d = 4\pi\int_{r_{\rm min}}^\infty dr\,r F_{\rm inc}(r),
\end{equation}
where the incident flux $F_{\rm inc}$ is given by
Eqs.~\ref{Finc_sphere} or~\ref{Finc_ellipsoid}.  This is the
luminosity intercepted by {\em both} sides of the disk.  ZLS define
the reflection fraction, averaged over all viewing angles, to be the
ratio of the luminosity intercepted by the disk and the thermal
Comptonization luminosity which can be seen by distant observers:
\begin{equation}
R_{\rm ZLS} = \frac{L_d}{L_x - L_d}.
\end{equation}
When comparing the observed $R$-$\Gamma$ correlation for a large
sample of objects which have random orientations, this is the
appropriate form to use for the reflection fraction.  However, for a
specific object with a known inclination, this expression may not be a
good approximation.  In particular, for an object with disk
inclination $i = 0$ and plasma radius $r_s \le \rmin$, the full X-ray
luminosity from the Comptonizing plasma is seen, but only half of the
disk reprocessing emission is observed.  In this case, one should use
\begin{equation}
R_{i=0} = \frac{L_d/2}{L_{x,{\rm app}}}.
\end{equation}
Note that the apparent X-ray luminosity appears in the denominator,
but $L_d$ in the numerator is computed using $L_x = \xi L_{x,{\rm
app}}$ to account for any anisotropy.  In Fig.~\ref{Gamma_vs_R}, we
plot the $R$-$\Gamma$ relationships using these two expressions for
the reflection fraction along with the fitted values from our analysis
of the {\em RXTE/PCA} data for NGC 3516 and from the analysis of
Nandra et al.\ (2000) for NGC 7469.  The model values using $R_{i=0}$
clearly under-predict the fitted values by a substantial amount and
are also much smaller than the average values, $R_{\rm ZLS}$.  As
Nandra et al.\ (2000) point out, any apparent correlation between
spectral index and Compton reflection strength may be partially due to
a statistical correlation between the parameters which arise from
model fitting rather than from intrinsic properties of the source.
The presence of such a statistical correlation between fitting
parameters suggests that the small reflection fractions given by this
model may be accommodated by the data if the underlying thermal
Comptonization spectra are actually somewhat harder than those found
by fitting the simple reflection models to the {\em RXTE/PCA} data.
Since {\em RXTE} observations generally do not provide much useful
spectral coverage above $\sim 15$\,keV for Seyfert galaxies, more
sensitive hard X-ray continuum measurements for energies $\ga 10$\,keV
are required to resolve this discrepancy.

Even in the absence of high quality hard X-ray data to measure the
thermal roll-over or the Compton reflection fraction definitively,
certain aspects of this model are indeed tested by the available data,
though the constraints are not as direct.  For example, if the UV
fluxes for NGC 7469 were consistent with an extrapolation of the $\nu
F_\nu \sim \nu^{4/3}$ dependence of the optical data, then a much
smaller disk inner truncation radius would be required.  This in turn
would imply a much broader Fe K$\alpha$ emission line that may
contradict the observations of a narrow line by Guainazzi et al.\
(1994), or it may require a much smaller black hole mass, which could
be in conflict with the reverberation mapping estimates of Collier et
al.\ (1998). Furthermore, the quantity $M\Mdot \sim
10^6$~\Msun$^2$~yr$^{-1}$ which we find for NGC 7469 (see
Table~\ref{7469_parameters}) is in rough accord with the value of
$M\Mdot \simeq 0.7 \times 10^6$~\Msun$^2$~yr$^{-1}$ found by Collier
et al.\ (1998) by fitting interband continuum lags for these same
observations.


The large disk inner truncation radius and vertically extended
Comptonizing region we find for NGC 7469 suggests that the inner
regions of this object may consist of an advection dominated accretion
flow (ADAF) similar to those which have been used to model the X-ray
spectra of Galactic X-ray binaries (e.g., Esin, McClintock, \& Narayan
1997; Esin et al.\ 2001).  However, the mass accretion rate which we
find for NGC 7469 in our optimal case, $\Mdot \simeq
0.15\,$\Msun~yr$^{-1}$, is larger than the Eddington limit by a factor
of $\sim 3$, assuming a Kerr metric and a $10^7$\,\Msun\ black hole;
and it is $\sim 40$\% of the Eddington limit for a standard thin disk
in a Schwarzschild metric.  This accretion rate likely exceeds the
critical accretion rate found by Narayan \& Yi (1995) of $\Mdot_{\rm
crit} \sim \alpha^2 \Mdot_{\rm Edd}$, below which advection dominated
flow is expected to occur.  The anomalous viscosity parameter is found
in magnetohydrodynamic simulations of disks to have values $\alpha \la
0.1$ (e.g., Balbus \& Hawley 1998).  As we noted above, we chose the
largest possible accretion rates for our models which were consistent
with a minimum plasma temperature of $k_BT \simeq 50$~keV and with the
variations in the optical and UV light curves being due to thermal
reprocessing.  We could have chosen smaller values of the accretion
rates in order to satisfy the limits imposed by $\Mdot_{\rm crit}$,
but that would necessitate increasing the plasma temperatures in order
to have larger thermal Comptonization luminosities, possibly violating
our implicit constraint of $kT \la m_e c^2$ for NGC 3516 and worsening
the situation for NGC 7469.  Furthermore, the lower accretion rates
would require much greater radiative efficiencies, and for NGC 3516,
the efficiency is already a fair fraction of unity.  We could trade
off reductions in the intrinsic disk luminosity with a larger
anisotropy factor, but values of $\xi \gg 1$ would likely be required.

However, for NGC 7469, our analysis really has only constrained the
quantity $M\Mdot \simeq 1.5 \times 10^6$~\Msun$^2$~yr$^{-1}$.  If the
central black hole mass were $M \simeq 10^8$~\Msun, then we would have
$\Mdot \simeq 0.03\,\Mdot_{\rm Edd}$ which would just satisfy the
accretion rate limit for advection for $\alpha \sim 0.2$, assuming a
Kerr metric.  Since the reverberation mapping estimates of the black
hole mass in NGC 7469 are given by $M \propto R_{\rm BLR} v_{\rm
BLR}^2/G$, where $R_{\rm BLR}$ and $v_{\rm BLR}$ are (model-dependent)
length and velocity scales which characterize the broad line region,
these quantities would have to be each underestimated by factors of
$\sim 10^{1/3} \sim 2$ in order to account for an
order-of-magnitude discrepancy in the central black hole mass (see
Peterson \& Wandel 2000).  For a Schwarzschild metric, a central black
hole mass of $M \simeq 3 \times 10^7$~\Msun would yield an accretion
rate $\Mdot \sim 0.04\Mdot_{\rm Edd}$.  In this case, $R_{\rm BLR}$
and $v_{\rm BLR}$ would need only each be underestimated by a factor
$\sim 1.4$.  For $M = 3 \times 10^7$~\Msun, we have $\rmin/r_g \sim
60$ which would still be consistent with the {\em inferred} appearance
of a narrow Fe K$\alpha$ line in the {\em RXTE/PCA} spectra.

By contrast, the required accretion rate for NGC 3516, assuming $M = 2
\times 10^7\,$\Msun\ and a Kerr metric, is $\Mdot \sim 0.09 \Mdot_{\rm
Edd}$.  This too nominally exceeds $\Mdot_{\rm crit}$; and since we
have assumed $\rmin = 6 r_g$, there is not much freedom here to
increase the central black hole mass in order to increase $\Mdot_{\rm
Edd}$ and decrease $\Mdot$.  If we assume $\rmin = 2 r_g$, which is
still consistent with the Fe K$\alpha$ emission line fits of Nandra et
al.\ (1999), the black hole mass could be larger by a factor $\sim 3$,
and we would have $\Mdot \sim 0.03 \Mdot_{\rm Edd}$.  In any case,
since the inner hot accretion flow we infer from our model is more
``disk-like'' than that of NGC 7469, a larger accretion rate relative
to $\Mdot_{\rm Edd}$ for NGC 3516 may nonetheless be consistent in
some sense with the ``strong ADAF proposal'' of Narayan \& Yi (1995;
see also Narayan, Mahadevan, \& Quataert 1998), which states that NGC
3516 should then be less likely to have an ADAF-like flow than NGC
7469.  Clearly, these machinations illustrate the speculative nature
and uncertainties involved in extracting physical significance from
our simple spectral fits based on the radiative processes.
Regardless, the more plausible of the above suggestions are not wildly
out of line with the observations or theoretical expectations.  This
perhaps indicates that this sort of modeling is on the cusp of
actually providing useful constraints on the dynamical models.

Numerous physical considerations which have not been included in these
calculations may affect our results.  In the case of a flattened
Comptonizing region, such as we found for NGC 3516, the mean Thomson
depth through the plasma will vary significantly depending on the
scattering order.  Along with the non-uniform illumination of the
Comptonizing region by the disk photons, this would lead us to expect
to see features in the X-ray spectrum similar to the anisotropy breaks
found by Stern et al.\ (1995) in their thermal Comptonization models
of disk coronae.  If the Comptonizing regions are threaded by
sufficiently strong magnetic fields, then thermal synchrotron
radiation can make a large contribution to the seed photon flux and
thus reduce the importance of the disk radiation in determining the
energy balance of the plasma as in the ADAF model of Esin et al.\
(2001).  However, Wardzi\'nski \& Zdziarski (2000) have argued that
for Seyfert 1s the disk thermal emission should be a much more
important source of seed photons than thermal synchrotron emission
except for very low luminosity sources which have $L/L_{\rm Edd} \la
10^{-4}$.  Finally, as we have already discussed, enhanced
illumination of the disk due to photons following curved geodesics or
anisotropy in the scattered radiation will clearly alter both the
thermally reprocessed disk flux as well as the Compton reflection
strength.

Despite these deficiencies, the relative success of this model in
reproducing the optical, UV and X-ray flux and spectral variability
observed from NGC 3516 and NGC 7469 underscores the crucial point that
{\em any} model which attempts to describe the emission in these
objects must at least take into account the roles that feedback and
energy balance play in determining the emission across a wide range of
energies.  As corollary to this, observations of these objects would
do well to consist of simultaneous broad band spectral coverage, from
the optical to the soft gamma-rays, in order to maximize the
constraints on this and future, more sophisticated radiative and
dynamical models.

\acknowledgements 

The author would like to thank the anonymous referee for very helpful
comments which have greatly improved this paper, Omer Blaes for many
useful discussions, and Mike Nowak for helpful comments on the
manuscript.  This work was partially supported by NASA ATP grant NAG
5-7723.  This research has made use of data obtained from the High
Energy Astrophysics Science Archive Research Center (HEASARC),
provided by NASA's Goddard Space Flight Center; and it has also made
use of the NASA/IPAC Extragalactic Database (NED) which is operated by
the Jet Propulsion Laboratory, California Institute of Technology,
under contract with NASA.

\clearpage

\appendix

\section{Formulae for Computing Disk Incident Fluxes and
Seed Photon Luminosities}

Deriving these formulae from the specified geometry is
straight-forward.  We present the following expressions for
completeness.  Here all lengths are in units of $\rmin$, i.e., $\rt =
r/\rmin$, etc..

\noindent
$\rt_s \ge 1$ (Spherical case, same as ZLS):
\begin{eqnarray}
\alpha_\max & = & \pi/2, \qquad \rt \le \rt_s, \\
            & = & \sin^{-1}(\rt_s/\rt), \qquad \rt > \rt_s,
\end{eqnarray}
\begin{eqnarray}
\phi_\max & = & \pi, \qquad \rt \le \rt_s, \\
              & = & \cos^{-1}\left[\frac{(\rt^2 - \rt_s^2)^{1/2}}
                    {\rt\cos\alpha}\right], \qquad \rt > \rt_s,
\end{eqnarray}
\begin{eqnarray}
\lt & = & \cos\alpha(\rt\cos\phi + [\rt_s^2(1 + \tan^2\alpha) 
                   - \rt^2(\sin^2\phi + \tan^2\alpha)]^{1/2}),
                   \qquad \rt \le \rt_s, \\
     & = & 2\cos\alpha[\rt_s^2 (1 + \tan^2\alpha) 
                   - \rt^2(\sin^2\phi + \tan^2\alpha)]^{1/2}, 
                   \qquad \rt > \rt_s.
\end{eqnarray}

\noindent
$\rt_s < 1$ (Ellipsoidal case):
\begin{equation}
\alpha_\max =  \tan^{-1}\left[\frac{\rt_s}{(\rt^2 - 1)^{1/2}}\right] \\
\end{equation}
\begin{equation}
\phi_\max  = \cos^{-1}\left[\frac{[(\rt^2 - 1)(\rt_s^2 
                                       + \tan^2\alpha)]^{1/2}}
                                     {\rt\rt_s}\right] \\
\end{equation}
\begin{equation}
\lt = \frac{2\rt_s[(\rt_s^2 + \tan^2\alpha) - \rt^2(\tan^2\alpha 
                          + \rt_s^2\sin^2\phi)]^{1/2}}
           {\cos\alpha(\rt_s^2 + \tan^2\alpha)}
\end{equation}

\clearpage

\clearpage

\begin{table}
\centering
\caption{Optical and X-ray data for the 1998 April Observations 
         of NGC 3516}
\smallskip
\label{3516_data}
\begin{tabular}{ccccccc}
\hline\hline
start time & duration & $F_{3590}$ & $F_{4235}$ & $F_{5510}$ 
   & $F_{2-10}$ & $\Gamma$ \\
(JD - 2450000.5) & (ks) 
   &\multicolumn{3}{c}{($10^{-14}$erg\,cm$^{-2}$s$^{-1}$\AA$^{-1}$)}
   & ($10^{-11}$erg\,cm$^{-2}$s$^{-1}$) & \\
\hline
 916.692 & 30.7 & 1.8052 & 0.9536 & 0.6294 & 6.75 & $1.693 \pm 0.024$\\
 917.048 & 44.2 & 1.8016 & 0.9518 & 0.6282 & 5.53 & $1.636 \pm 0.015$\\
 917.559 & 46.2 & 1.8182 & 0.9606 & 0.6340 & 6.29 & $1.671 \pm 0.018$\\
 918.094 & 25.9 & 1.8271 & 0.9652 & 0.6371 & 5.81 & $1.660 \pm 0.020$\\
 918.393 & 31.6 & 1.8278 & 0.9656 & 0.6373 & 6.03 & $1.657 \pm 0.017$\\
 918.759 & 74.6 & 1.7945 & 0.9480 & 0.6257 & 4.94 & $1.628 \pm 0.015$\\
\hline
\end{tabular}
\end{table}

\begin{table}
\centering
\caption{Fixed Model Parameters, Luminosities, and Radiative
Efficiencies for NGC 3516}
\smallskip
\label{3516_parameters}
\begin{tabular}{lcccc}
\hline\hline
parameter   & \multicolumn{3}{c}{value} \\
\hline
$\cos i$              & \multicolumn{3}{c}{0.820} \\
$d_l$ ($10^{26}$\,cm) & \multicolumn{3}{c}{1.22} \\
$M$ ($10^7$\,\Msun)   & 1 & 2 & 3                \\
$\rmin$ ($10^{13}$\,cm) & 0.89 & 1.78 & 2.67     \\
$\Mdot$ ($10^{-2}$\Msun yr$^{-1}$) & 1.86 & 0.946 & 0.633 \\
$L_{\rm untr}$ ($10^{44}$\,erg s$^{-1}$)    & 5.26 & 2.67 & 1.79 \\
$L_\disk$ ($10^{44}$\,erg s$^{-1}$)         & 1.90 & 0.97 & 0.65 \\
$L_{x,{\rm avg}}$ ($10^{44}$\,erg s$^{-1}$) & 1.05 & 0.75 & 0.72 \\
$ (L_{x,{\rm avg}} + L_\disk)/\Mdot c^2$    & 0.28 & 0.32 & 0.38 \\
$(L_{x,{\rm avg}} + L_\disk)/L_{\rm Edd}$   & 0.23 & 0.07 & 0.03 \\
\hline
\end{tabular}
\end{table}

\begin{table}
\centering
\caption{Fixed Model Parameters, Luminosities, and Radiative
Efficiencies for NGC 7469}
\smallskip
\label{7469_parameters}
\begin{tabular}{lcccc}
\hline\hline
parameter   & \multicolumn{3}{c}{value} \\
\hline
$\cos i$              & \multicolumn{3}{c}{0.866} \\
$d_l$ ($10^{26}$\,cm) & \multicolumn{3}{c}{2.06} \\
$M$ ($10^7$\,\Msun)   & \multicolumn{3}{c}{1} \\
$\rmin$ ($10^{13}$\,cm) & \multicolumn{3}{c}{30.} \\
$\xi$ (anisotropy) & 1 & 1.5 & 1.5 \\
$\Gamma$ (allowed variation) & \multicolumn{2}{c}{fixed} & $\pm 1\sigma$ \\
$\Mdot$ (\Msun yr$^{-1}$) & 0.159 & 0.116 & 0.150 \\
$L_{\rm untr}$ ($10^{44}$\,erg s$^{-1}$)    & 45.0 & 32.8 & 42.5 \\
$L_\disk$ ($10^{44}$\,erg s$^{-1}$)         & 0.63 & 0.46 & 0.60 \\
$L_{x,{\rm avg}}$ ($10^{44}$\,erg s$^{-1}$) & 1.43 & 1.93 & 1.47 \\
$ (L_{x,{\rm avg}} + L_\disk)/\Mdot c^2$    & 0.023 & 0.036 & 0.024 \\
$(L_{x,{\rm avg}} + L_\disk)/L_{\rm Edd}$   & 0.16 & 0.19 & 0.16  \\
\hline
\end{tabular}
\end{table}

\clearpage

\end{document}